# Emergent half-metal at finite temperatures in a Mott insulator

Gour Jana[1], Abhishek Joshi[1], Subhajyoti Pal[1] & Anamitra Mukherjee [1 ✉]

Sustaining exotic quantum mechanical phases at high temperatures is a long-standing goal of condensed matter physics. Among them, half-metals are spin-polarized conductors that are essential for realizing room-temperature spin current sources. However, typical half-metals are low-temperature phases whose spin polarization rapidly deteriorates with temperature increase. Here, we first show that a low-temperature insulator with an unequal charge gap for the two spin channels can arise from competing Mott and band insulating tendencies. We establish that thermal fluctuations can drive this insulator to a half-metal through a first-order phase transition by closing the charge gap for one spin channel. This half-metal has 100% spin polarization at the onset temperature of metallization. Further, varying the strength of electron repulsion can enhance the onset temperature while preserving spin polarization. We outline experimental scenarios for realizing this tunable finite temperature half-metal.

[1] School of Physical Sciences, National Institute of Science, Education and Research, HBNI, Jatni 752050, India. ✉email: anamitra@niser.ac.in

Spin-polarized metals are of great technological importance, having applications in diverse fields ranging from spintronic devices to quantum computation[1,2]. Thus material realizations of half-metals continue to be a significant research direction on both theoretical[3–12] and experimental fronts[13–15]. Candidates for half-metals such as double perovskites[4–7], Heusler alloys[8–10,13–15], and manganites[16,17], have therefore been extensively investigated, among others. Introducing vacancies in transition metal oxides such as NiO, MnO[18], fluorinated BN, ZnO[19], and ferrimagnetic inverse spinels such as $Fe_3S_4$[20] have also been proposed as systems that can host half-metallicity. However, the degradation of spin polarization with temperature increase and experimental challenges in material growth[21] has long been barriers to producing spin current sources at room temperature in all half-metal candidates. Thus, predictions of half-metallic states in simple models are vital for uncovering novel mechanisms for half-metallicity and subsequent material realizations that can potentially circumvent the aforementioned difficulties.

In this context, we consider the ionic Hubbard model (IHM), with local repulsion $U$ and staggered onsite "ionic" potential $\Delta$, introduced initially for studying charge transfer effects in organic crystals[22] and for modeling ferroelectric perovskites[23]. In the half-filled IHM for $U = 0$, a finite $\Delta$ opens up a band-gap that promotes large occupation of the sub-lattice with $-\Delta$ potential. In the opposite limit, for $\Delta = 0$ and large $U$, singly occupied sites are preferred on both sub-lattices. The simultaneous presence of the two agencies leads to a competition that causes a zero temperature band insulator to Mott insulator transition with increasing $U$, for a fixed $\Delta$. This band to Mott insulator transition has been demonstrated in one dimension[24–28] and using dynamical mean-field theory (DMFT) on square[29,30] and Bethe lattice[29,31,32]. More importantly, the half-filled model in one dimension[33] and on square lattice using DMFT[34] have been shown to support zero temperature half-metallic state on introducing next nearest neighbor (nnn) hopping. However, transport response and survival of spin polarization of this half-metal, in presence of thermal fluctuations, has remained largely unaddressed.

In this article, by computing spin-dependent transport response at finite temperatures, we first establish a low-temperature antiferromagnetic half-metal (HM$_1$), bracketed by a paramagnetic band insulator for $U < U_B$ and an anti-ferromagnetic Mott insulator for $U > U_M$. The spin polarization of HM$_1$ is maximized as $T \to 0$. Next, we reveal a ferrimagnetic half-metal (HM$_2$) that emerges out of thermal excitations of the Mott insulating state for $U$ values in the vicinity of $U_M$. We show that, unlike HM$_1$, the Mott insulating ground state ensures that half-metallicity in HM$_2$ occurs only above an onset temperature via a thermally driven first-order insulator to metal transition. We demonstrate that at the onset temperature of metallization, HM$_2$ is fully spin-polarized. We find that the onset temperature is highly sensitive to correlation strength, acting as a control knob for tuning the temperature window of half-metallicity. Raising the onset temperature does not degrade the spin polarization in its vicinity. We explain the origin of HM$_2$ and map out the parameter space of its stability. We close with a discussion on the experimental realization of HM$_2$ in correlated oxides and cold atomic systems.

We define the IHM on a square lattice as follows,

$$H = -t \sum_{\langle i,j \rangle, \sigma} (c_{i\sigma}^\dagger c_{j\sigma} + h.c) - t' \sum_{\langle\langle i,j \rangle\rangle, \sigma} (c_{i\sigma}^\dagger c_{j\sigma} + h.c) \\ + \Delta \sum_{i \in A} n_i - \Delta \sum_{i \in B} n_i + U \sum_i n_{i\uparrow} n_{i\downarrow} - \mu \sum_i n_i \quad (1)$$

where, $c_{i\sigma}^\dagger$ ($c_{i\sigma}$) are electron creation (annihilation) operators at the site $i$ with spin $\sigma$. $t$ and $t'$ are respectively the nearest and nnn hopping amplitudes. We choose $t'/t < 0$ in our study. $n_{i\sigma} = c_{i\sigma}^\dagger c_{i\sigma}$ is the number operator for spin $\sigma$ at a site $i$ and $n_i$ is the spin summed local number operator. $\Delta$ is the magnitude of "ionic" potential and takes positive (negative) values on the $A$ ($B$) sub-lattice. $U$ is the local Hubbard repulsion. $\mu$ denotes the chemical potential, and is adjusted to maintain half-filling. We employ a recently developed semiclassical Monte Carlo approximation (s-MC)[35–37] to investigate the finite temperature properties of IHM on large lattice sizes.

"Methods" sub-sections titled, Treatment of the interaction term, Solution strategy, and The nature of approximation cover the (s-MC) scheme in detail. In particular, in the "Methods"—The nature of approximation, it is discussed that s-MC is numerically inexpensive, can capture results beyond finite temperature mean-field theory, and has a reasonable quantitative agreement with Determinantal Quantum Monte Carlo (DQMC) over a wide temperature range[35,37–39].

## Results

**Transport response regimes.** In Fig. 1, we show the spin-resolved resistivity, $\rho_\sigma(T)$, extracted from the low-frequency optical conductivity, as discussed in the "Methods"—Optical conductivity. We show the case of low $U$ band insulator in Fig. 1a, followed by two intermediate $U$ examples in Fig. 1b and Fig. 1c that, as will be detailed below, support HM$_1$ and HM$_2$ respectively. Finally in Fig. 1d we provide an example of the large $U$ Mott insulator. In all the plots, triangles and squares denote $\rho_\uparrow(T)$ and $\rho_\downarrow(T)$ respectively. We first compare the small $U$ band insulator in Fig. 1a with the large $U$ Mott insulator in Fig. 1d. In Fig. 1a, we find typical insulating nature ($d\rho_\sigma/dT < 0$) for both spin channels. We can see in Fig. 1a that $d\rho/dT$ remains negative for all temperatures. This is because the correlation strength is insufficient to screen the one body potential and close the charge gap as temperature increases. Similar behavior was found in DQMC study of IHM for the $t' = 0$[40]. For the chosen values of $\Delta$ and $t'$, this band insulating response occurs for $U < U_B (=2.8t)$, while the Mott insulator lies beyond $U_M (=4.75t)$. We see typical Mott insulating behavior representative of large $U(=5.5t)$ in Fig. 1d with diverging resistivity as $T \to 0$. Further, we see that the sign of $d\rho_\sigma/dT$ for both spin channels changes from negative to positive around $T/t \sim 0.12$, signaling a crossover from an insulator to a metal, mediated by thermal fluctuations. This temperature scale is the analog of $T^*$ in the standard half-filled Hubbard model, with zero $\Delta$ and $t'$, that governs the crossover from a paramagnetic insulator to a paramagnetic metal, previously seen in DQMC[41] and s-MC[35].

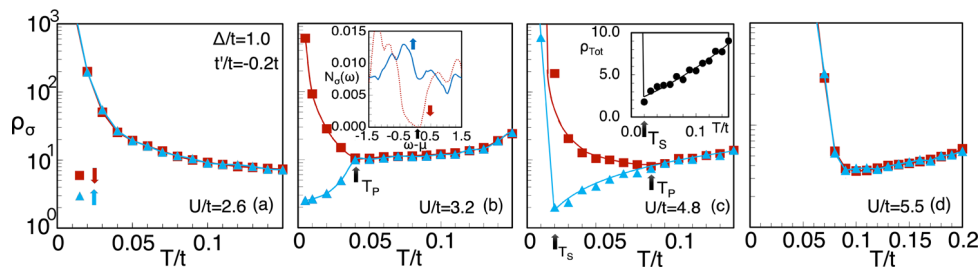

**Fig. 1 Temperature evolution of spin-dependent transport. a–d** Show spin-resolved resistivity, $\rho_\sigma(T)$ for correlation strength $U$ values as indicated. $\rho_\sigma(T)$ is calculated in units of $(\pi e^2 / \hbar a_0)$, with $e$, $\hbar$ and $a_0$ being the electronic charge, the Planck's constant divided by $2\pi$ and the lattice spacing respectively. The next nearest neighbor hopping $t'/t$ and charge transfer energy $\Delta/t$ are chosen to be $-0.2$ and $1.0$, respectively. The squares denote $\rho_\downarrow$ and triangles represent $\rho_\uparrow$. The inset in **b** shows the low $T(= 0.005t)$, spin-resolved density of states (DOS), $N_\sigma(\omega)$ for $U/t = 3.2$, with a small arrow at the bottom indicating the Fermi energy. The up and down spin channels are also indicated by arrows. The inset in **c** shows the total resistivity $\rho(T)$ for $U/t = 4.8$. **d** $\rho_\sigma(T)$ for the robust Mott insulator at $U/t = 5.5$. Results are shown for $32^2$ system size with periodic boundary conditions. Standard deviation of the low-frequency optical conductivity data is used to estimate error bars in the resistivity plots. Symbol sizes used are larger than the error bars.

In contrast to these limiting cases, in the regime $U_B < U < U_M$ and $U_M < U < 5t$, we find distinctly different thermal evolution of resistivity. Representative data for these two regimes are shown in Fig. 1b, c respectively. In Fig. 1b, spin resolved resistivity for $U/t = 3.2$ shows that $d\rho_\downarrow/dT$ is negative for $T < 0.04t$, while $d\rho_\uparrow/dT$ exhibits metallic behavior for all temperatures. At $T = 0.04t$, the of sign $d\rho_\downarrow/dT$ changes and $\rho_\uparrow$ becoming equal to $\rho_\downarrow$. We define $T/t = 0.04$ as $T_P$, the lowest temperature where $\rho_\uparrow$ and $\rho_\downarrow$ become identical. We will later show that below $T_P$, we have a spin-polarized metal, whose polarization approaches 100% as $T \to 0$. We refer to this half-metallic regime as $HM_1$ and $T_P$ define a depolarization temperature scale, above which spin polarization is completely lost.

In the resistivity data shown in Fig. 1c for $U/t = 4.8$, we find diverging $\rho_\sigma$ as $T \to 0$, as is expected in a Mott state. However, at $T/t = 0.02$, $d\rho_\uparrow/dT$ switches sign while $d\rho_\downarrow/dT$ continues to remain negative up to $T/t = 0.09$. From the inset in Fig. 1c, showing the total resistivity as a function of temperature, we identify $T/t = 0.02$ as the onset of metallic response, where $d\rho_\uparrow/dT$ had changed its sign. We denote this onset temperature by $T_S$. Also as for $HM_1$, we denote $T/t = 0.09$ as $T_P$, the temperature above which $\rho_\uparrow(T) = \rho_\downarrow(T)$. Thus, for $U \gtrsim U_M$, we find a window in temperature, $T_S < T < T_P$, where the Mott insulating ground state gives way to a finite-temperature half-metal, $HM_2$. We will show that, within numerical accuracy, $HM_2$ is fully spin polarized near the onset temperature $T_S$.

**Density of states and spin polarization**. The inset in Fig. 1b shows the spin-resolved density of states (DOS), $N_\sigma(\omega)$, defined in the "Methods"—The spin-resolved DOS. The result is shown for $U/t = 3.2$ and at the lowest temperature of our calculation ($T/t = 0.005$). We find that only one "up" spin channel contributes to the weight at Fermi energy for very low temperature, as would be expected for usual half-metals. In contrast, low-temperature DOS for $U/t = 4.8$ in Fig. 2a shows a finite charge gap that is unequal for the two spin channels. Consequently, we find that the temperature-induced filling of the charge gap is much more rapid for the "up" than for the "down" spin channel. We see this behavior in Fig. 2b, which shows the temperature evolution of $N_\sigma(0)$, the spin-resolved spectral weight extracted from the DOS, at $\omega = \mu$. The filling of the charge gap starts at $T_S (=0.02t)$ in a "spin asymmetric" manner, and the asymmetry persists up to $T_P (=0.09t)$. We discuss the origin of this spin-asymmetric filling of the charge gap later in the paper. We add that the Mott state at low temperature is always characterized by an unequal charge gap for the spin channels for all $U$ values investigated in our study. However, beyond $U/t = 5$, the temperature-induced filling of the charge gap occurs equally for both spin channels as seen in Supplementary Fig. 1a presented in Supplementary Note 1, and prohibits the formation of the $HM_2$ phase. The low-temperature s-MC DOS qualitatively agrees with DMFT calculations at $T = 0$[34].

To quantify the spin polarization of the half-metals at finite temperature, we define the transport spin polarization as $P(T) = \langle J_\uparrow^x - J_\downarrow^x \rangle / \langle J_\uparrow^x + J_\downarrow^x \rangle$, where $J_\sigma^x$ is the spin resolved current operator along the $x$-direction. Here and later in the paper, the angular brackets indicate quantum as well as thermal averaging. Details of the calculation are provided in "Methods"—Optical conductivity. Fig. 2c, shows $P(T)$ for $U/t = 3.2$, 4.8 and 4.9. For $HM_1$ at $U/t = 3.2$, we find 100% spin polarization or $P(T) \sim 1$ at $T = 0.005t$. With temperature increase, the "down" spin channel that was gapped in the ground state, e.g. the inset of Fig. 1b, starts to fill up, causing the polarization to decay, eventually suppressing it to zero beyond $T_P(=0.04t)$. $P(T)$ for $HM_2$, at $U/t = 4.8$, is shown by circles. We find that $P(T) = 0$ in the insulating regime, for $T < T_S(=0.02t)$. It then abruptly jumps to 1 at $T_S$, the onset of half-metallicity. With temperature increase beyond $T_S$, $P(T)$ drops until it reaches zero at $T_P = 0.09t$, coinciding with the closing of the window of asymmetric charge gap filling, as seen in Fig. 2b. We also see that while $T_S$ increases from $0.02t$ for $U = 4.8t$ to $0.06t$ for $U = 4.9t$, $P(T)$ at $T_S$, remains unchanged. We thus find an unexpected property that $T_S$ enhancement does not sacrifice spin polarization for $HM_2$. We note that at $T_S$, for all parameter points of $HM_2$, the insulating to metallic spin channels resistivity ratio is $\sim 10^2 - 10^3$. This translates into $P(T) \sim 1$ or 100% spin polarization at $T_S$, as can be seen from the expression of $P(T)$ in the "Methods"—Optical conductivity.

**Ferrimagnetic half-metal at finite temperature**. To define magnetic order at finite temperature in two dimensions, we add a small SU(2) symmetry breaking magnetic field, as described in the "Methods"—The treatment of the interaction term. In Fig. 3a, we show the thermal evolution of the average sub-lattice magnetizations, $S_\alpha^z \equiv (\langle n_{\alpha\uparrow} \rangle - \langle n_{\alpha\downarrow} \rangle)/2$, $\alpha \in \{A, B\}$ for $U/t = 4.8$. We present the formulae for spin and sub-lattice resolved densities in "Methods"—The spin and sub-lattice resolved density. For $T < T_S$, we find that $S_A^z = -S_B^z$, indicating long-range antiferromagnetic order. However, for $T_S < T < T_P$, we see that $|S_A^z| > |S_B^z|$, while the staggered nature of magnetic order remains unchanged.

In Fig. 3b, we show the temperature evolution of the difference between the average sub-lattice local moments ($\delta M = M_A - M_B$), where $M_\alpha = \langle n_\alpha \rangle - 2 \langle n_{\alpha\uparrow} n_{\alpha\downarrow} \rangle$, and $\alpha \in [A, B]$. We find that $\delta M = 0$ for $T < T_S$, but has a finite value in the window $T_S < T < T_P$. Fig. 3c shows the net spin polarization of the system for $U/t = 4.8$ with triangles, $\delta n = \langle n_\uparrow \rangle - \langle n_\downarrow \rangle$. The correlation between these $\delta M$ and $\delta n$ is apparent from Fig. 3b and c. The temperature dependence of the sub-lattice magnetizations in Fig. 3a follows from the relation $S_A^z + S_B^z = \delta n/2$. Due to the unequal magnitude of the staggered sub-lattice magnetizations, we refer to $HM_2$ as a *ferrimagnetic* half-metal. We also observe that, as a generic feature of $HM_2$, $\delta n$, $\delta M$, and $P(T)$ undergo an abrupt jump at $T = T_S$, as seen here for $U/t = 4.8$. This behavior strongly indicates a thermal fluctuation-induced "first-order" transition from an antiferromagnetic Mott insulator to a ferrimagnetic half-metal. In Supplementary Fig. 1b reported in Supplementary Note 2, we provide typical hysteresis behavior for $\delta M$ associated with the first-order transition, obtained through a cooling and heating protocol for $HM_2$. Since our calculation uses fixed temperature steps that only bracket the actual critical temperature for the insulator to metal transition, we refer to the temperature scale of metallization as an "onset" scale. We find no discontinuity in the temperature evolution of sub-lattice magnetization for $HM_1$ as seen in Supplementary Fig. 1c presented in Supplementary Note 3.

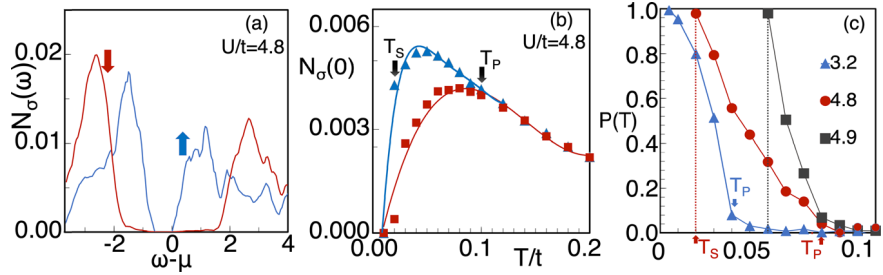

**Fig. 2 Thermal evolution of density of states (DOS) & polarization. a** This shows the low T density of states, $N_\sigma(\omega)$ for $U/t = 4.8$. The large arrows denote the spin channels. **b** It shows thermal evolution of $N_\sigma(0)$, the spectral weight at the chemical potential for $U/t = 4.8$. Here triangles denote $N_\uparrow(0)$ and squares indicate $N_\downarrow(0)$. The locations of the onset temperature of half metallicity $T_S$ and depolarization temperature $T_P$ are marked with black arrows. **c** Shows spin polarization $P(T)$ as defined in Section 2 of the main text titled "Density of states and spin polarization", for indicated $U$ values. Various $T_S$ and $T_P$ are demarcated with arrows color coded with the symbols. Lines are a guide to the eye. All results are shown for $t'/t = -0.2$ and $\Delta/t = 1.0$. Error bars in **b** and **c** are computed from the standard deviation in the spin-resolved DOS, and low-frequency optical conductivity data respectively. Symbol sizes used are larger than the error bars.

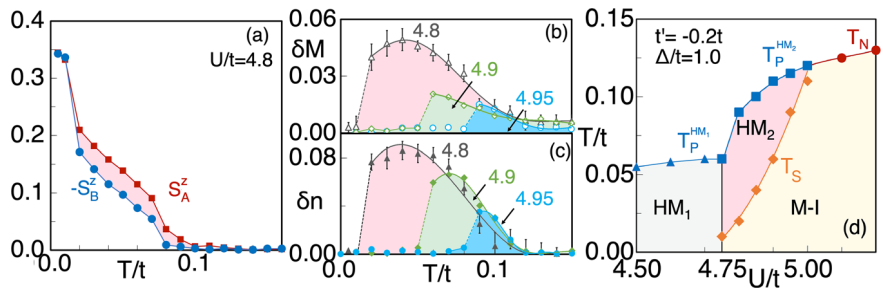

**Fig. 3 Magnetism & $U - T$ phase diagram. a** It shows the sub-lattice magnetizations $S_A^z$ and $S_B^z$ for $U/t = 4.8$. **b, c** Show the difference in sub-lattice local moments ($\delta M$) and the total spin polarization of the system ($\delta n$) respectively, for $U/t = 4.8$ 4.9 and 4.95. In each case, the dashed lines between two temperature points indicate the location of the onset temperature for half-metallicity. **d** This shows the $U - T$ phase diagram with a low-temperature antiferromagnetic half-metal HM$_1$ to antiferromagnetic Mott insulator (M-I) transition at $U_M = 4.75t$. Triangles indicate $T_P$, the depolarization temperature of half metallicity for HM$_1$. The red shaded region refers to the finite-temperature ferrimagnetic half-metal HM$_2$ for $U_M < U < 5t$ and bounded by the onset temperature of polarization $T_S$ and the depolarization temperature $T_P$. The $T_S$ ($T_P$) scales for HM$_2$ are demarcated by diamonds (squares). Circles refer to $T_N$, the temperature above which the M-I state loses long-range antiferromagnetic order. All results are for $t'/t = -0.2$, $\Delta/t = 1.0$ and on $32^2$ system. Standard deviations in the sub-lattice magnetization in **a** and the various ordering temperature scales in **d** are smaller than the symbol sizes; Error bars for $U = 4.8t$ are generated from standard deviations the sub-lattice local moments in **b** and spin-resolved densities in **c**. Error bars for $U = 4.9t$ and $U = 4.95t$ are similar in magnitude to that for $U = 4.8t$ in **b** and **c**.

To understand the above temperature-driven behavior, we first note that finite $\Delta$ causes a double occupation penalty of $U + 2\Delta$ ($U - 2\Delta$) on the A (B) sub-lattice in the zero hopping limit. This differential penalty persists for finite hopping, albeit with a renormalized value. Supplementary Fig. 1d in Supplementary Note 4 shows typical temperature evolution of the sub-lattice double occupations $\langle n_{\alpha\uparrow}n_{\alpha\downarrow} \rangle$ with $\alpha \in (A, B)$ for HM$_2$ for $U = 4.8t$. It clearly shows that, $\langle n_{B\uparrow}n_{B\downarrow} \rangle > \langle n_{A\uparrow}n_{A\downarrow} \rangle$ at all temperatures. In Fig. 3a, we see that the larger cost of double occupation at A, initially triggers a finite magnetization $S_A^z$ at $T_P(= 0.09t)$, while $S_B^z$ becomes non-zero only at a lower temperature. It also maintains $|S_B^z| < |S_A^z|$ and consequently a finite $\delta n$ following $S_A^z + S_B^z = \delta n/2$, below $T_P$ and down to $T_S$. The sub-lattice local moments difference $\delta M$ in Fig. 3b, also follows from this differential double occupation penalty. However, energetically, the large build-up of $\langle n_{B\uparrow}n_{B\downarrow} \rangle$ becomes untenable at low temperatures. Below $T_S$, the difference in the sub-lattice double occupations is reduced through a first-order transition, making the sub-lattice magnetizations equal, which quenches $\delta n$ and limits the half-metallic state HM$_2$ to finite temperature. From Fig. 1c, we find that the conduction electron spin direction is identical to the net spin polarization $\delta n$, 'up' in this instance. It also translates into $N_\uparrow(0) > N_\downarrow(0)$ for $T_S < T < T_P$, and can be interpreted as thermal fluctuation driven spin-asymmetric filling of charge gap, as seen in Fig. 2b.

**U-T phase diagram**. We now discuss the $U - T$ phase diagram, Fig. 3d, to investigate the stability of HM$_2$ and the tunability of $T_S$. For $T/t \leq 0.02$, $U_M(=4.75t)$ separates HM$_1$ and M-I phases. Figure 3d is a part of the full phase diagram shown in Supplementary Fig. 2 and discussed in detail in Supplementary Note 5. As seen in Supplementary Fig. 2, HM$_1$ starts at $U = 2.8t$ and the $T_P$ (triangles) increases with $U$, reaching a maximum of ($0.055t$) close to $U_M$. From static magnetic structure factor discussed in the "Methods"—Static magnetic structure factor, we find the HM$_1$ has an antiferromagnetic background and the magnetic transition temperature $T_N$, coincides with $T_P$. Supplementary Fig. 1c discussed in Supplementary Note 3 depicts typical temperature dependence of the sub-lattice magnetizations for HM$_1$. The HM$_2$ phase emerges from the M-I in the range $U_M < U < 5t$ and above a $U$-dependent $T_S$, indicated by diamonds. The ferrimagnetic order in HM$_2$ is destabilized at the corresponding $T_P$ values (squares), as seen for example, in Fig. 3a for $U = 4.8t$. Circles represent the antiferromagnetic transition temperature for the M-I phase beyond the HM$_2$ window or $U > 5t$.

We also see that $T_S$ increases rapidly with $U$ for HM$_2$. This behavior follows from the need of greater thermal energy to close the Mott gap at larger $U$ values. However, as $U$ grows, we also move deeper into the Mott state, progressively suppressing the effect of $\Delta$. The increasingly dominant role of $U$ manifests as a systematic reduction in magnitudes of $\delta M$ and $\delta n$ as is seen in Fig. 3b, c. In these plots, we compare the temperature profiles of $\delta M$ and $\delta n$ for three $U$ values. We find that while with $U$ increase, the window of finite $\delta M$ and $\delta n$ move to higher temperatures, their magnitudes are systematically suppressed. As the sub-lattice moment magnitudes approach each other with increasing $U$, $T_P$ and $T_S$ merge into a single transition temperature $T_N$, of the (M-I) phase beyond $U = 5t$. Finally, from Fig. 3d, we see that $T_S$ increases by 350% over the window in $U$ for which HM$_2$ is stable. In the discussion section, we propose 4d transition metal element-based double perovskites as potential candidates for realizing HM$_2$. We obtain a realistic estimate of the $T_S$ and $T_P$ using a hopping scale of 0.2–0.3 eV for double perovskite. It yields $T_S$ to be about 70–100 K and $T_P$ about 230–350 K in the middle of HM$_2$ in Fig. 3d at $U/t = 4.875$.

In concluding this section, we would like to briefly discuss the role of $t'$ in stabilizing HM$_2$. $t'$ acts as a source of frustration that inhibits formation of long-range antiferromagnetic order and promotes metallic tendency. In fact, both half-metallic phases are absent for $|t'/t| < 0.05$. Increase in the magnitude of $t'/t$ beyond 0.05, systematically shifts $U_M$ to larger $U$ values to overcome frustration effects and consequently increases $T_S$ as seen in Supplementary Fig. 3a discussed in Supplementary Note 6. However, once $U_M$ gets too big for a large enough $|t'/t|$, the reduced impact of $\Delta$ is inadequate to support HM$_2$, i.e., the temperature-driven gap-filling becomes spin symmetric, as previously described. We find that HM$_2$ is no longer realized for $|t'/t| \geq 0.3$, at $\Delta/t = 1$. In Supplementary Fig. 3b, presented in Supplementary Note 6, we show preliminary results from an exhaustive numerical search, from which we infer that HM$_2$ can be stabilized for Mott insulators with $\Delta/U \sim 0.2$–0.3, over a large window of $U$ and $\Delta$. Thus, HM$_2$ has a broad stability window in the $U - \Delta - t'$ parameter space, and Fig. 3d depicts only a fixed $\Delta - t'$ cross-section of this regime.

## Discussion
Unlike all previous proposals where half metallicity is a ground state property, here we have demonstrated that a half-metal can emerge from a Mott insulating ground state. The insulating ground state ensures that the half-metal with 100% spin polarization occurs at a finite temperature and protects against temperature-induced depolarization effects. The correlation strength-dependent enhancement of the onset temperature naturally carries the fully spin-polarized metal to higher temperatures without depolarizing. We want to emphasize that the low-temperature half-metal HM$_1$, has negligible $\delta n$ and zero $T_S$ due to the reduced effect of $U$ and, in this respect, is different from HM$_2$. Lastly, HM$_2$ does not require interaction between large Hund's coupled core spins and itinerant electrons like in the double exchange mechanism; it is different from the Slater–Pauling rule-based, low-temperature half-metallic Heusler alloys and $T = 0$ half-metallicity in doped Mott insulators away from half-filling[42]. The experimental realization of this unique finite-temperature half-metal will be a significant step toward the goal of creating spin-polarized current sources at room temperature. With this in mind, we close with a discussion on the possible realization of HM$_2$ in solid-state and cold atomic experimental setups.

For correlated oxides representing the IHM, we consider epitaxial thin films of cubic double perovskites, X$_2$ABO$_6$. Here, A and B represent two species of 4d transition metal (TM) atoms, and X can be rare-earth, alkaline-earth, or alkali elements. To prevent high $U$ and Hund's coupling effects, we do not consider 3d TM elements. In addition, we avoid 5d TM elements with strong spin-orbit coupling. In the 4d TM series, {Zr, Nb, Mo, Tc, Ru and Rh} starting with Zr, difference in the charge transfer energy ($\Delta$) between successive elements range between 0.2 eV and 0.7 eV[43]. Also the local correlation strength $U$ is moderate, ranging between 1eV and 3eV[44,45]. Hence, $\Delta/U \sim 0.3$ required for (HM$_2$), can be achieved. In an octahedral crystal field, the 4d orbitals are split by 3–4 eV[46], into high energy $e_g$ and low energy $t_{2g}$ levels. We expect that (A, B) combinations of 4d TM elements with a small $\Delta$ and relatively large crystal field splitting will facilitate the formation of partially filled $t_{2g}$ bands. Suitably chosen X site element can achieve half-filled $t_{2g}$ manifold and also can tune $U$ by controlling the electronic bandwidth. Finally, nnn hopping is relevant for 4d TM elements[47,48]. Hence one can potentially identify a number of candidates that are insulators with small charge gap, such as Sr$_2$RuMoO$_6$[49]. Based on the above-mentioned literature we can crudely estimate $U_{Re/Mo} \sim 3$ eV, $\Delta \sim 1$–2 eV and a $t_{2g} - e_g$ splitting of 3 eV. These estimates yield $\Delta/U \sim 0.3$–0.6 eV, which is within the ballpark of the ratio needed for HM$_2$ for Sr$_2$RuMoO$_6$. We expect complete quenching of orbital angular momentum in a half-filled $t_{2g}$ 4d orbital; however, one has to investigate the role of small Hund's coupling[50], prior to identifying realistic material candidates.

While the model has not yet been realized on a square lattice in the cold atomic setup, the experimental techniques for such a realization exist. For example IHM has already been realized for fermionic cold atomic systems for hexagonal lattice[51]. This experiment investigated the band to Mott insulator transition by tracking suppression of double occupation over wide variations of $\Delta$ and $U$ [0, 41$t$] and [0, 30$t$], respectively, in the units of nearest-neighbor hopping strength $t/h \sim 174$ Hz. This range of parameter variation should be easily possible for the square lattice as well. $t'/t$ ratio can be tuned over a large window on shaken optical lattices[52]. Also, the highest $T_S/t \sim 0.1$ is within the ballpark of the experimental temperature scales of $0.2t$[53,54]. We suggest that, $\delta n$ would be the natural quantity to measure as a signature of HM$_2$, analogous to spin polarization measurements for metallic (Stoner) ferromagnets in cold atomic systems[53].

## Methods
We briefly present the formal derivation of the s-MC approximation scheme, calculation method, nature of the approximation, and definitions of observables.

**Treatment of the interaction term.** To set up the semiclassical Monte-Carlo s-MC method, we first decouple the interaction term in Eq. (1) using standard Hubbard–Stratonovich (H–S) transformation, by introducing auxiliary fields (Aux. F.). For this we express the interaction term at each lattice site as the sum of square of the local number operator $n_i$ and the local spin operator $S_i$ as follows:

$$U n_{i\uparrow} n_{i\downarrow} = U\left[\frac{1}{4}(n_i)^2 - (\mathbf{S_i} \cdot \hat{\Omega})^2\right] \quad (2)$$

here, the spin operator at the $i$th site (running over A and B sub-lattices) is $\mathbf{S_i} = \frac{1}{2}\sum_{\alpha\beta} c_{i\alpha}^\dagger \boldsymbol{\sigma}_{\alpha\beta} c_{i\beta}$, with $\hbar \equiv 1$ and $\hat{\Omega}$ is an arbitrary unit vector. The partition function for the IHM Hamiltonian, $H \equiv H_0 + H_1$ is $Z = Tr e^{-\beta H}$ with $H_0$ and $H_1$ containing the one body and interaction terms respectively. The trace is taken over the occupation number basis. $\beta = 1/T$, is the inverse temperature where $k_B$ is set equal to 1. The window $[0, \beta]$ is divided into $M$ equally spaced slices separated by interval $\Delta\tau$, with $\beta = M\Delta\tau$. The slices are labeled by $l$. In the limit $\Delta\tau \to 0$ by using Suzuki-Trotter decomposition, we write $e^{-\beta(H_0+H_1)} \approx (e^{-\Delta\tau H_0} e^{-\Delta\tau H_1})^M$ to the first order in $\Delta\tau$. For a given imaginary time slice '$l$', the partition function for the interaction term can then be written as,

$$const. \times \int d\phi_i(l) d\xi_i(l) \times e^{-\Delta\tau\left[\sum_i \left(\frac{\phi_i(l)^2}{U} + i\phi_i(l)n_i + \frac{\xi_i(l)^2}{U} - 2\xi_i(l)\hat{\Omega}\cdot\mathbf{S_i}\right)\right]} \quad (3)$$

Here, we have introduced (Aux. F.)'s $\phi_i(l)$ and $\xi_i(l)$ at each site and imaginary time slice. Following literature[55], we combine the unit vector $\hat{\Omega}$ and $\xi_i(l)$ as a new vector auxiliary field $\mathbf{m}_i(l)$, defined as $\xi_i(l)\hat{\Omega}$. $\phi_i(l)$ couples to the local charge density operator and $\mathbf{m}_i(l)$ couples to the local spin operator. Thus the full partition function is proportional to,

$$Tr \prod_{l=M}^{1} \int d\phi_i(l) d\mathbf{m}_i(l) \times e^{-\Delta\tau\left[H_0 + \sum_i \left(\frac{\phi_i(l)^2}{U} + i\phi_i(l)n_i + \frac{\mathbf{m}_i(l)^2}{U} - 2\mathbf{m}_i(l)\cdot\mathbf{S_i}\right)\right]} \quad (4)$$

In the above, the integrals are taken over $\{\phi_i(l), \mathbf{m}_i(l)\}$, and the order of product for the slices run from $l=M$ to the $l=1$. We note that at this stage the partition function is exact, manifestly SU(2) symmetric and the $\{\phi_i(l), \mathbf{m}_i(l)\}$ fields fluctuate in space and imaginary time. We now make the approximation of (i) retaining only the spatial dependence of the (Aux. F.) variables and (ii) use a saddle point value of the (Aux. F.) $\phi_i(l)$ equal to $iU\langle n_i\rangle/2$. This allows us to extract the following effective spin-fermion type Hamiltonian $H_{eff}$ where fermions couple to *classical* (Aux. F.) $\{\mathbf{m}_i\}$. For further details, we refer to our earlier work[35].

$$H_{eff} = H_o + \frac{U}{2}\sum_i\left[(\langle n_i\rangle n_i - 2\mathbf{m}_i\cdot\mathbf{S}_i) + \frac{1}{2}(\mathbf{m}_i^2 - \langle n_i\rangle^2)\right] \quad (5)$$

In $H_0$, we include a small SU(2) symmetry-breaking magnetic field term. The field avoids well-known Mermin-Wagner issues and allows for an unambiguous definition of long-range magnetic order in two dimensions at finite temperature. It also provides a global axis for the definition of *up* and *down* spin components. We have checked that the spin polarization in the band insulator and the HM$_1$ phases are zero within numerical accuracy, establishing that the small symmetry breaking term does not induce spurious spin polarizations. We solve the spin-fermion model by well-established cluster-based exact diagonalization coupled with classical Monte-Carlo (ED+MC) method[36,56]. Below we briefly outline the main steps of the ED+MC and refer to our earlier work detailing the solution methodology[35].

**Solution strategy**. We start the calculation at high temperature $T/t=1$ with random $\{\mathbf{m}_i\}$ and $\{\langle n_i\rangle\}$ fields at each site. The $\{\langle n_i\rangle\}$ fields define the $\{\phi_i\}$ variables, as mentioned above. We then sample the $\{\mathbf{m}_i\}$ fields by sequentially visiting all lattice sites. For this, we first diagonalize $H_{eff}$ for the chosen (Aux. F.) configuration and then propose an update at a site. To decide the acceptance of a proposed update for $\mathbf{m}_i$ at the $i$th site, we employ a standard Metropolis scheme. In this approach, the usual Boltzmann factor governs the acceptance probability, which depends on the energy from the exact diagonalization before and after the proposed update. This update process done sequentially at all lattice sites constitutes a Monte-Carlo system sweep. We conduct 6000 sweeps at each temperature. After every 20 sweeps, we perform a self-consistency loop to converge the $\{\langle n_i\rangle\}$ fields for a fixed $\{\mathbf{m}_i\}$ configuration. We leave the first 2000 sweeps for equilibration and compute observables from the remaining 4000 sweeps, skipping every five sweeps to avoid self-correlation effects. We reduce the temperature in small steps and repeat the above protocol at each temperature. Finally, we average all observables over 20 separate runs with independent random initial choices of $\{\mathbf{m}_i\}$ and $\{\langle n_i\rangle\}$ configurations at the starting (high) temperature.

**Nature of the approximation**. Earlier work[35] has established that at low temperatures, the Monte Carlo sampling leads to uniform $\mathbf{m}_i$ configurations akin to unrestricted Hartree-Fock results. However, at higher temperatures where thermal fluctuations begin to dominate quantum fluctuations, the thermal sampling of the (Aux. F.)'s $\{\mathbf{m}_i\}$ captures temperature effects considerably more accurately than a simple finite temperature mean-field theory, making s-MC a progressively superior approximation. s-MC based study of the half-filled Hubbard model in two and three dimensions[35,36], and with nnn hopping[37], have revealed that this approach can capture several results that are beyond simple finite temperature mean-field theory. These include non-monotonic dependence of antiferromagnetic ordering scale on $U$, finite $T$ paramagnetic insulating phase with local moments, pseudo-gap to normal-metal crossover, and specific heat systematics with temperature. Moreover, these results have a reasonable quantitative agreement with DQMC[38,39]. s-MC is free of the usual fermion sign problem, analytical continuation issues and is numerically inexpensive. Recent s-MC based studies include finite-temperature properties of the Anderson-Hubbard model[57], and temperature-driven Mott transition in frustrated triangular-lattice Hubbard model[58].

**Observable definitions**. Here we define the various observables used in the paper.

**Optical conductivity**. The total d.c conductivity $\sigma_{dc}$ along the $x$-direction, is computed from the Kubo-Greenwood formula[59] for optical conductivity.

$$\sigma(\omega) = \frac{\pi e^2}{Nha}\sum_{\alpha,\beta}(n_\alpha - n_\beta)\frac{|f_{\alpha\beta}|^2}{\epsilon_\beta - \epsilon_\alpha}\delta(\omega - (\epsilon_\beta - \epsilon_\alpha)) \quad (6)$$

$f_{\alpha\beta}$ are the matrix elements for the current operator. The explicit form of $f_{\alpha\beta}$ is $\langle\psi_\alpha|J_x|\psi_\beta\rangle$. The current operator is given by $J_x = -ia_0\sum_{i,\sigma}[t(c^\dagger_{i,\sigma}c_{i+a_0\hat{x},\sigma} - h.c) + t'(c^\dagger_{i,\sigma}c_{i+a_0\hat{x}+a_0\hat{y},\sigma} - h.c)]$. Here, $|\psi_\alpha\rangle$ and $\epsilon_\alpha$ are single-particle eigenstates and associated eigenvalues respectively. $n_\alpha = f(\mu - \epsilon_\alpha)$ is the Fermi function. We calculate the average d.c. conductivity, $\sigma_{dc}$, by integrating over small frequency window, $\sigma_{dc} = (\Delta\omega)^{-1}\int_0^{\Delta\omega}\sigma(\omega)d\omega$. The interval is chosen to be $\Delta\omega \sim 0.005t$. For the spin-resolved conductivity, $\sigma_{dc,\sigma}$ we use appropriate spin resolved states and operators to construct $f^\sigma_{\alpha\beta} = \langle\psi_\alpha|J^\sigma_x|\psi_\beta\rangle$ in the conductivity expression. The resistivity is obtained for the inverse of average dc conductivity. Finally, $P(T)$ is calculated from $\frac{\sigma_{dc,\uparrow}(\mu) - \sigma_{dc,\downarrow}(\mu)}{\sigma_{dc,\uparrow}(\mu) + \sigma_{dc,\downarrow}(\mu)}$ using $J^\sigma_x \propto \sigma_{dc,\sigma}$, with the convention that the 'up' refers to the spin channel of the electrons that delocalizes to form the half-metal.

**The spin resolved DOS**. The spin-resolved DOS is defined as follows: $N_\sigma(\omega) = \sum_{\gamma\alpha}\langle\alpha,\sigma|\psi_\gamma\rangle|^2\delta(\omega - \epsilon_\gamma)$, where $\epsilon_\gamma$ and $|\psi_\gamma\rangle$ are the eigenvalues and eigenvectors of $H_{eff}$. Here $\alpha \in \{A, B\}$, the two sub-lattices and $\sigma$ refers to spin. Lorentzian representation of the above $\delta$-function is used to compute DOS. The broadening of the Lorentzian is ~$BW/2N$ with $BW$ being the non-interacting bandwidth and $N$ is the total number of lattice sites.

**The spin and sub-lattice resolved density**. The average spin and sub-lattice resolved density is given by

$$\langle n_{\alpha\sigma}\rangle = \frac{2}{N}\sum_{i\in\alpha,\gamma}|\langle i\sigma|\psi_\gamma\rangle|^2 f(\epsilon_\gamma - \mu) \quad (7)$$

where, $i$ is the site index. $\alpha \in A, B$, and $\sigma$ is the spin index. $\epsilon_\gamma, |\psi_\gamma\rangle$ and $f(\epsilon_\lambda - \mu)$ are as defined above. $N$ is the total number of lattice sites. The sub-lattice resolved occupation is calculated by summing over densities for $A$ and $B$, for each spin channel independently. The magnetization in Fig. 3a are constructed from these densities.

**Static magnetic structure factor**. $S_q$ as defined below, is computed for $\mathbf{q}=(\pi,\pi)$, from which $T_N$ is extracted for the various phases.

$$S_q = \frac{1}{N}\sum_{i,j}e^{i\mathbf{q}\cdot(\mathbf{r}_i-\mathbf{r}_j)}\langle\mathbf{S}_i\cdot\mathbf{S}_j\rangle \quad (8)$$

All symbols have the usual meaning as defined above and $i, j$ run over all lattice sites.

**Data availability**
The data that support the findings of this study are available from the corresponding author upon request.

**Code availability**
Codes used to produce the findings of this study are available from the corresponding author upon reasonable request.

Received: 24 August 2021; Accepted: 1 March 2022;
Published online: 21 March 2022

## References
1. Zutic, I., Fabian, J. & Das Sarma, S. Spintronics: fundamentals and applications. *Rev. Mod. Phys.* **76**, 323–410 (2004).
2. Li, X. & Yang, J. First-principles design of spintronics materials. *Natl Sci. Rev.* **3**, 365–381 (2016).
3. Katsnelson, M. I., Irkhin, V. Y., Chioncel, L., Lichtenstein, A. I. & de Groot, R. A. Half-metallic ferromagnets: from band structure to many-body effects. *Rev. Mod. Phys.* **80**, 315–378 (2008).
4. Chen, S. H., Xiao, Z. R., Liu, Y. P. & Wang, Y. K. Investigation of possible half-metallic antiferromagnets on double perovskites LaABB′O$_6$ (A=Ca,Sr,Ba; B,B′ =transition elements). *J. Appl. Phys.* **108**, 093908 (2010).
5. Wang, Y. K., Lee, P. H. & Guo, G. Y. Half-metallic antiferromagnetic nature of La$_2$VTcO$_6$ and La$_2$VCuO$_6$ from ab-initio calculations. *Phys. Rev. B* **80**, 224418 (2009).
6. Erten, O. et al. Theory of half-metallic ferrimagnetism in double perovskites. *Phys. Rev. Lett.* **107**, 257201 (2011).
7. Xu, J. et al. Prediction of room-temperature half-metallicity in layered halide double perovskites. *npj Comput. Mater.* **5**, 114 (2019).
8. Kundu, A., Ghosh, S., Banerjee, R., Ghosh, S. & Sanyal, B. New quaternary half-metallic ferromagnets with large curie temperatures. *Sci. Rep.* **7**, 1803 (2017).
9. Ren, Y., Cheng, F., Zhang, Z. H. & Zhou, G. Half metal phase in the zigzag phosphorene nanoribbon. *Sci. Rep.* **8**, 2932 (2018).
10. Chaudhuri, S. et al. Half metallicity in Cr substituted Fe$_2$TiSn. *Sci. Rep.* **11**, 524 (2021).
11. Hashmi, A., Nakanishi, K., Farooq, M. U. & Ono, T. Ising ferromagnetism and robust half-metallicity in two-dimensional honeycomb-kagome Cr$_2$O$_3$ layer. *npj 2D Mater. Appl.* **4**, 39 (2020).
12. Nie, Y.-m & Hu, X. Possible half-metallic antiferromagnet in a hole-doped perovskite cuprate predicted by first-principles calculations. *Phys. Rev. Lett.* **100**, 117203 (2008).
13. Müller, G. M. et al. Spin polarization in half-metals probed by femtosecond spin excitation. *Nat. Mater.* **8**, 56–61 (2009).
14. Jourdan, M. et al. Direct observation of half-metallicity in the Heusler compound Co$_2$MnSi. *Nat. Commun.* **5**, 3974 (2014).

15. Battiato, M. et al. Distinctive picosecond spin polarization dynamics in bulk half metals. *Phys. Rev. Lett.* **121**, 077205 (2018).
16. Osborne, I. S. Half-metallic manganites. *Science* **299**, 627–627 (2003).
17. Dagotto, E. Nanoscale Phase Separation and Colossal Magnetoresistance. The Physics of Manganites and Related Compounds (2013).
18. Ködderitzsch, D., Hergert, W., Szotek, Z. & Temmerman, W. M. Vacancy-induced half-metallicity in MnO and NiO. *Phys. Rev. B* **68**, 125114 (2003).
19. Kan, E. J. et al. Prediction for room-temperature half-metallic ferromagnetism in the half-fluorinated single layers of BN and ZnO. *Appl. Phys. Lett.* **97**, 122503 (2010).
20. Wu, M., Zhou, X., Huang, S., Cheng, J. & Ding, Z. A first-principles study of the effect of vacancy defects on the electronic structures of greigite ($Fe_3S_4$). *Sci. Rep.* **8**, 11408 (2018).
21. Prophet, S., Dalal, R., Kharel, P. R. & Lukashev, P. V. Half-metallic surfaces in thin-film $Ti_2MnAl_{0.5}Sn0.5$. *J. Phys. Condens. Matter* **31**, 055801 (2018).
22. Nagaosa, N. & Takimoto, J.-i Theory of neutral-ionic transition in organic crystals. I. Monte Carlo simulation of modified Hubbard model. *J. Phys. Soc. Jpn.* **55**, 2735–2744 (1986).
23. Egami, T., Ishihara, S. & Tachiki, M. Lattice effect of strong electron correlation: implication for ferroelectricity and superconductivity. *Science* **261**, 1307–1310 (1993).
24. Resta, R. & Sorella, S. Many-body effects on polarization and dynamical charges in a partly covalent polar insulator. *Phys. Rev. Lett.* **74**, 4738–4741 (1995).
25. Wilkens, T. & Martin, R. M. Quantum Monte Carlo study of the one-dimensional ionic Hubbard model. *Phys. Rev. B* **63**, 235108 (2001).
26. Gidopoulos, N., Sorella, S. & Tosatti, E. Born effective charge reversal and metallic threshold state at a band insulator-Mott insulator transition. *Eur. Phys. J. B - Condens. Matter Complex Syst.* **14**, 217–226 (2000).
27. Kampf, A. P., Sekania, M., Japaridze, G. I. & Brune, P. Nature of the insulating phases in the half-filled ionic Hubbard model. *J. Phys.: Condens. Matter* **15**, 5895–5907 (2003).
28. Go, A. & Jeon, G. S. Phase transitions and spectral properties of the ionic Hubbard model in one dimension. *Phys. Rev. B* **84**, 195102 (2011).
29. Garg, A., Krishnamurthy, H. R. & Randeria, M. Can correlations drive a band insulator metallic? *Phys. Rev. Lett.* **97**, 046403 (2006).
30. Kancharla, S. S. & Dagotto, E. Correlated insulated phase suggests bond order between band and Mott insulators in two dimensions. *Phys. Rev. Lett.* **98**, 016402 (2007).
31. Bag, S., Garg, A. & Krishnamurthy, H. R. Phase diagram of the half-filled ionic Hubbard model. *Phys. Rev. B* **91**, 235108 (2015).
32. Craco, L., Lombardo, P., Hayn, R., Japaridze, G. I. & Müller-Hartmann, E. Electronic phase transitions in the half-filled ionic Hubbard model. *Phys. Rev. B* **78**, 075121 (2008).
33. Japaridze, G. I., Hayn, R., Lombardo, P. & Müller-Hartmann, E. Band-insulator–metal–mott-insulator transition in the half-filled $t-t'$ ionic hubbard chain. *Phys. Rev. B* **75**, 245122 (2007).
34. Bag, S., Garg, A. & Krishnamurthy, H. R. Correlation driven metallic and half-metallic phases in a band insulator. *Phys. Rev. B* **103**, 155132 (2021).
35. Mukherjee, A. et al. Testing the Monte Carlo-Mean field approximation in the one-band Hubbard model. *Phys. Rev. B* **90**, 205133 (2014).
36. Mukherjee, A., Patel, N. D., Bishop, C. & Dagotto, E. Parallelized traveling cluster approximation to study numerically spin-fermion models on large lattices. *Phys. Rev. E* **91**, 063303 (2015).
37. Jana, G. & Mukherjee, A. Frustration effects at finite temperature in the half filled Hubbard model. *J. Phys.: Condens. Matter* **32**, 365602 (2020).
38. Paiva, T., Scalettar, R. T., Huscroft, C. & McMahan, A. K. Signatures of spin and charge energy scales in the local moment and specific heat of the half-filled two-dimensional Hubbard model. *Phys. Rev. B* **63**, 125116 (2001).
39. Paiva, T., Scalettar, R., Randeria, M. & Trivedi, N. Fermions in 2D optical lattices: temperature and entropy scales for observing antiferromagnetism and superfluidity. *Phys. Rev. Lett.* **104**, 066406 (2010).
40. Paris, N., Bouadim, K., Hebert, F., Batrouni, G. G. & Scalettar, R. T. Quantum Monte Carlo study of an interaction driven band insulator to metal transition. *Phys. Rev. Lett.* **98**, 046403 (2007).
41. Rost, D., Gorelik, E. V., Assaad, F. & Blümer, N. Momentum-dependent pseudogaps in the half-filled two-dimensional Hubbard model. *Phys. Rev. B* **86**, 155109 (2012).
42. Garg, A., Krishnamurthy, H. R. & Randeria, M. Doping a correlated band insulator: a new route to half-metallic behavior. *Phys. Rev. Lett.* **112**, 106406 (2014).
43. Zhong, Z. & Hansmann, P. Band alignment and charge transfer in complex oxide interfaces. *Phys. Rev. X* **7**, 011023 (2017).
44. Somia et al. First-principles study of perovskite molybdates $AMoO_3$ (A = Ca, Sr, Ba). *J. Electron. Mater.* **48**, 1730–1739 (2019).
45. M. Musa Saad H.-E., First-principles DFT study new series of ruthenates double perovskites $Ba_2MRuO_6$ with M = Sc, Ti, V, Cr, Mn, Fe and Co. *Mater. Chem. Phys.* **204**, 350–360 (2018).
46. Takayama, T., Chaloupka, J., Smerald, A., Khaliullin, G. & Takagi, H. Spin-orbit-entangled electronic phases in 4d and 5d transition-metal compounds. *J. Phys. Soc. Jpn.* **90**, 062001 (2021).
47. Mustonen, O. et al. Magnetic interactions in the S = 1/2 square-lattice antiferromagnets $Ba_2CuTeO_6$ and $Ba_2CuWO_6$: parent phases of a possible spin liquid. *Chem. Commun.* **55**, 1132–1135 (2019).
48. Sarma, D. D., Mahadevan, P., Saha-Dasgupta, T., Ray, S. & Kumar, A. Electronic structure of $Sr_2FeMoO_6$. *Phys. Rev. Lett.* **85**, 2549–2552 (2000).
49. Agiorgousis, M. L., Sun, Y., Choe, D., West, D. & Zhang, S. Machine learning augmented discovery of chalcogenide double perovskites for photovoltaics. *Adv. Theory Simul.* **2**, 1800173 (2019).
50. Georges, A., Medici, L. D. & Mravlje, J. Strong correlations from Hund's coupling. *Annu. Rev. Condens. Matter Phys.* **4**, 137–178 (2013).
51. Messer, M. et al. Exploring competing density order in the ionic Hubbard model with ultracold fermions. *Phys. Rev. Lett.* **115**, 115303 (2015).
52. Struck, J. et al. Tunable gauge potential for neutral and spinless particles in driven optical lattices. *Phys. Rev. Lett.* **108**, 225304 (2012).
53. Greif, D., Uehlinger, T., Jotzu, G., Tarruell, L. & Esslinger, T. Short-range quantum magnetism of ultracold fermions in an optical lattice. *Science* **340**, 1307–1310 (2013).
54. Mazurenko, A. et al. A cold-atom Fermi-Hubbard antiferromagnet. *Nature* **545**, 462–466 (2017).
55. Schulz, H. J. Effective action for strongly correlated fermions from functional integrals. *Phys. Rev. Lett.* **65**, 2462–2465 (1990).
56. Kumar, S. & Majumdar, P. A travelling cluster approximation for lattice fermions strongly coupled to classical degrees of freedom. *Eur. Phys. J. B - Condens. Matter Complex Syst.* **50**, 571–579 (2006).
57. Patel, N. D., Mukherjee, A., Kaushal, N., Moreo, A. & Dagotto, E. Non-fermi liquid behavior and continuously tunable resistivity exponents in the Anderson-Hubbard model at finite temperature. *Phys. Rev. Lett.* **119**, 086601 (2017).
58. Tiwari, R. & Majumdar, P. Spectroscopic signatures of the Mott transition on the anisotropic triangular lattice. *Europhys. Lett.* **108**, 27007 (2014).
59. Mahan, G. Many Particle Physics (Kluwer Academic, 1958).

### Acknowledgements
We acknowledge the use of the NOETHER and VIRGO clusters at NISER for numerical computations and fruitful discussion with V. Ravi Chandra, Ashok Mohapatra, Ashis Kumar Nandy, and Nitin Kaushal.

### Author contributions
G.J., A.J., and S.P. contributed equally. A.M. supervised this work.

### Competing interests
The authors declare no competing interests.

### Additional information
**Supplementary information** The online version contains supplementary material available at https://doi.org/10.1038/s42005-022-00847-w.

**Correspondence** and requests for materials should be addressed to Anamitra Mukherjee.

**Peer review information** *Communications Physics* thanks Biplab Sanyal and the other, anonymous, reviewer(s) for their contribution to the peer review of this work. Peer reviewer reports are available.

**Reprints and permission information** is available at http://www.nature.com/reprints

**Publisher's note** Springer Nature remains neutral with regard to jurisdictional claims in published maps and institutional affiliations.